# Prescribed Pattern Transformation in Swelling Gel Tubes by Elastic Instability

Howon Lee,[1] Jiaping Zhang,[2] Hanqing Jiang,[2] and Nicholas X. Fang[1],*

[1]Department of Mechanical Engineering, Massachusetts Institute of Technology,

Cambridge, MA 02139, USA

[2]School for Engineering of Matter, Transport, and Energy, Arizona State University,

Tempe, AZ 85287, USA

We present a study on swelling-induced circumferential buckling of tubular shaped gels. Inhomogeneous stress develops as gel swells under mechanical constraints, which gives rise to spontaneous buckling instability without external force. Full control over the post-buckling pattern is experimentally demonstrated. A simple analytical model is developed using elastic energy to predict stability and post-buckling patterns upon swelling. Analysis reveals that height to diameter ratio is the most critical design parameter to determine buckling pattern, which agrees well with experimental and numerical results.

46.32.+x, 62.20.mq, 81.40.Jj

Buckling instability has been studied extensively for the past few decades as one of the most critical structural failure modes [1]. This classical theme is recently gaining new attention as a useful way for creation and transformation of patterns because buckling is often accompanied with large deformation and radical shape change of the structure. Nature has already developed such techniques to leverage mechanical instability to create a wealth of complex patterns. As biological tissues and organisms grow non-uniformly or under constraints, plane features transform into rich patterns with complexity as found in such examples as wavy edges of plant leaves [2], fine annular patterns in fingerprints [3, 4], and inter-connected creases of brain cortex [5].

This elegant approach to achieve pattern transformation by harnessing mechanical instability has not been much explored until recent progress in material science and manufacturing technologies for soft materials such as elastomers and hydrogels. Particularly, swelling gels have attracted increasing interest because they can actively grow and shrink depending on environmental conditions such as humidity, temperature, and pH [6-8]. Hydrogel-based structures, therefore, can spontaneously create and reversibly pose different patterns via buckling without the need for external load as opposed to classical structural buckling where mechanical force is essential to trigger mechanical instability. This holds great potential in development of self-operating devices with switchable functionalities.

Recently, simple geometries of hydrogel systems have been investigated in this context, including hydrogel layers with different swelling ratio [9, 10], gel strips and disks bound to a hard substrate [11-13], and polymer films with periodic pores [14]. Taking advantage of buckling for pattern formation, however, has not been extended to more diverse shapes and geometries to further explore inspirations from nature where hosts of dynamic features are found. Of particular interest in our study is tubular structure with circumferential wrinkles, which is not only geometrically interesting by itself, but also holds crucial physiological significance in biomedical engineering. For example, wrinkled airway of asthmatic bronchiole results from the swollen inner cell layer. Hence, its topology and corresponding mechanical condition are essential factors to better understand how diseased cells behave and interact with the physiological circumstances [15, 16]. Although numerical and theoretical studies on non-planar geometries such as a tube and a sphere have been reported [17-19], few have been able to create soft tubular structures with well controlled dimensions and constraints due primarily to lack of three-dimensional (3D) fabrication technology for hydrogels. To reproduce complex patterns emerging in such structures has been even more challenging.

Here we present demonstration of well-controlled pattern transformation of micro-structured tubular gels using swelling-induced circumferential buckling. Principle behind the pattern formation is explained by simple energy analysis and design criteria to control instability pattern is presented. We employed a novel 3D micro-fabrication technology, projection micro-stereolithography (PμSL) [20], to fabricate hydrogel micro tubes, the bottom of which is fixed to impose constraints against swelling as shown in Fig. 1(a). Characteristic dimensional parameters, *t*, *h*, and *D* represent thickness, height, and diameter of the model structure in dry state, respectively. Subjected to the fixed boundary condition on the bottom, the gel develops inhomogeneous stress when allowed to swell. By appropriate selection of dimension, constrained swelling can be made to exhibit buckling instability, causing the circular wall to transform into wrinkled patterns with different wave numbers.

To demonstrate pattern formation, we fabricated tubular gel samples in different dimensions using poly(ethylene glycol) diacrylate (PEGDA) hydrogel (see Supplemental Material [21]). Four groups of samples (**I-IV**) with different levels of normalized thickness *t/h* were prepared, with group **I** being thicker and group **IV** being more slender.

Each group consists of six samples (**i-vi**) with different levels of normalized height *h/D*, with the sample **i** being shorter and the sample **vi** being taller (see Supplemental Material for physical dimensions of each sample [21]). For swelling experiment, we placed a sample upside down and put in the bath with water covered with oil layer on top as illustrated in Fig. 1(a). Then the sample was brought into contact with water surface for swelling, while base substrate part on which the gel tube was fixed stayed in the top oil layer. In this way, water can diffuse into the tube wall allowing the sample swell before the constraining base relaxes by wetting. Circular tubes transformed into a wide variety of rich patterns as swelling proceeded, and the evolution of the gel morphology was recorded by a charge-coupled device (CCD) camera throughout the course of the experiment (see Movie S1 and Movie S2).

Figure 1(b) presents swelling patterns obtained in the swelling experiment from different samples. Result of the experiment suggests strong connection between the normalized thickness *t/h* and stability, and between the normalized wall height *h/D* and buckling pattern. Samples in group **I** and **II** had tendency to remain stable during swelling, while samples in group **III** and **IV** underwent mechanical instability and transformed into wrinkled patterns. More interestingly, samples with the same normalized height *h/D* transformed into instability patterns with the number of wrinkles close to each other, regardless of the normalized wall thickness *t/h*. The same trend was observed when we repeated the same experiment using hydrogel with different stiffness and swelling ratio, confirming that dimension plays a dominant role in spontaneous buckling in swelling gel. To better understand and generalize this observation, morphology of swelling gel tube is studied via energy analysis. Cylindrical coordinate system (*r*, *θ*, *z*) is used to describe deformation of the model system. The interface between tubes and substrates have been treated to achieve very strong bonding in the experiment, which would otherwise influence buckling formation and mode [22]. Therefore, base of the tube is considered as a fixed boundary condition, whereas the rest of the structure is free to swell. When allowed to swell, there are two possible configurations for the swollen tube to adopt in order to accommodate expanded geometry in the original dimension; compression and buckling, as illustrated in Fig. 1(a). The system chooses a shape that minimizes total potential energy. We assume that the entire structure first swells to the fully swollen state and then the bottom end is forced to fit into its original dry dimension. Therefore, fully swollen state is considered as stress-free reference state, and any deformation from it has an elevated potential energy of the system. Although swelling of gel under constraints involves a coupling of elastic and chemical equilibrium, this coupling effect is negligible because the free energy due to the coupling is assumed to be a constant for both compression and buckling configurations, permitting pure elastic consideration. For compressed configuration in which the structure stays stable, the cross-section of the tube remains circular with the radius varying with height. Therefore, increase in potential energy results solely from the in-plane compression of the structure. On the other hand, situation is not as simple in buckled configuration where wrinkled wall undergoes bending in both circumferential and axial direction. We analyze the elastic energy for both configurations to predict stability as well as post-buckling pattern (see Supplemental Material [21]).

For stable configuration, cross-section of the tube remains circular with the radius being a function of height only. We assume that radius is linearly varying from the dry radius, $R$

(=***D/2***), at the fixed bottom to the fully swollen radius, ***λR***, at the top as shown in Fig. 2(a), where ***λ*** denotes length-wise equilibrium swelling ratio. Since only in-plane compression is involved, total elastic energy in the stable configuration is obtained as

$$U_{stable} = \int \frac{1}{2} E \varepsilon_{\theta\theta}^2 dV = \frac{1}{24} \pi E D t h \cdot b(\lambda), \qquad (1)$$

where $E$ is Young's modulus of the fully swollen gel and $b(\lambda) = (1 - 1/\lambda)^2 (3 + \lambda)$.

Once the structure becomes mechanically unstable, it buckles and creates wrinkles along its circumference. In this configuration, the swollen gel fits into confined geometry by posing a wavy shape at cost of elastic energy as shown in Fig. 2(b). There are two parts of energy involved in this case. The first part is the elastic energy due to the wavy bending along the circumferential direction. This energy contribution increases with buckling mode because the wall undergoes more bending with large curvature in higher buckling mode. The second part is the elastic energy due to the deflection of the wall in axial direction. As the gel swells more near its free upper end than near its confined bottom, the gel wall has to deflect outwards or inwards in axial direction depending on the position on the wave. This energy contribution decreases with buckling mode because higher buckling mode results in smaller wave amplitude in given length, thus less deflection in axial direction is necessary. Therefore, with the two energy contributions working together, there exists an optimum buckling mode that gives minimum total potential energy.

This argument can be formulated by assuming sinusoidal wave pattern with amplitude obtained in a closed form using approximation for the elliptic integral [23]. Radius of wrinkled cross-section can be written as

$$r(\theta,z) = R\left\{1 + \frac{2}{n}a(\lambda)\cos n\theta \cdot \frac{z^2(3h-z)}{2h^2}\right\}, \tag{2}$$

where $n$ is a number of waves along the circumference (i.e., buckling mode) and $a(\lambda) = [(2/(3-\lambda))^2 - 1]^{1/2}$ (See Supplemental Material [21]). The elastic energy along the circumferential direction is integrated from the local bending energy with varying curvature determined by the sinusoidal wave,

$$U_{bending} = \frac{1}{2}\int EI\kappa^2 \, dV = \frac{11}{140}\frac{\pi E t^3 a^2(\lambda) h}{D}n^2, \tag{3}$$

where $I = \frac{t^3 R d\theta}{12}$ and $\kappa \approx \frac{1}{R^2}\frac{\partial^2 r(\theta,z)}{\partial \theta^2}$ are bending moment of inertia and curvature of the wave, respectively.

The energy due to the deflection away or towards the central axis is modeled as bending of a set of cantilever beams surrounding the central axis as shown in Fig. 2(b). For simplicity, equivalent springs for cantilever beams are introduced. Then each spring undergoes stretching by the distance to the neutral circumferential line, which is the wave amplitude at each point, $\Delta r = r(\theta; z=h) - R$. From beam theory, equivalent spring constant of each cantilever beam is given by

$$k_{eq} = \frac{E t^3 R d\theta}{4h^3}. \tag{4}$$

Energy for the wall is obtained by integration of energy of each individual cantilever beam:

$$U_{conf} = \int \frac{1}{2}k_{eq}(\Delta r)^2 = \frac{\pi E t^3 D^3 a^2(\lambda)}{16 h^3} \cdot \frac{1}{n^2}. \tag{5}$$

Combining Eq. (3) and Eq. (5), total elastic energy for buckled configuration is obtained as an analytical form given by

$$U_{unstable} = \frac{\pi E t^3 h a^2(\lambda)}{D} \left[ \frac{11}{140} n^2 + \frac{1}{16(h/D)^4} \cdot \frac{1}{n^2} \right]. \tag{6}$$

It is interesting to find that two terms are proportional to $n^2$ and $1/n^2$, respectively. The former is from circumferential bending (lower energy for lower mode) and the latter is from axial deflection (lower energy for higher mode). We find that this opposite dependence on mode number from two energy contributions involved in (6) brings the system to a certain buckling mode in the event of buckling. Moreover, it is surprising to see that dimensional parameters involved in this competition in the bracket in (6) are **h** and **D** only. In other words, other seemingly-important parameters such as **t**, **λ**, and **E** have no impact on the determination of buckling mode. This trend is verified from our experiments shown in Fig. 1(b). Once the tube buckles, the buckling mode does not depend on **t**, but only on **h/D**.

Figure 3(a) plots the total potential energy for different possible buckling modes as a function of **h/D**. We can clearly see that for each **h/D**, there is a mode number $\tilde{n}$ which brings the potential energy to the minimum, suggesting corresponding buckling patterns for given dimension. Taking $\partial U_{unstable}/\partial n = 0$ yields

$$\tilde{n} = \left(\frac{35}{44}\right)^{\frac{1}{4}} \cdot \frac{1}{(h/D)} = \frac{0.944}{(h/D)}. \tag{7}$$

As the actual buckling mode number **n** is an integer, its value is either $n = \lfloor \tilde{n} \rfloor$ or $n = \lceil \tilde{n} \rceil$, depending on which one gives lower potential energy. "$\lfloor \ \rfloor$" and "$\lceil \ \rceil$" represent the floor and ceiling functions which map a real number to its largest previous or smallest following integer, respectively.

Between stable and buckled state, the system chooses the configuration at the lower energy level. Instability index is defined as follows from (1) and (6) to characterize relative magnitude of the energy levels,

$$\gamma = \frac{U_{stable}}{U_{unstable}}\bigg|_{n=\tilde{n}} = \frac{1}{(t/h)^2} \cdot c(\lambda), \tag{8}$$

where $c(\lambda) = (\sqrt{35/11}/6)[b(\lambda)/a^2(\lambda)]$ is a swelling factor increasing monotonically with $\lambda$. $\gamma > 1$ means $U_{stable} > U_{unstable}$, thus the system opts to buckle, while $\gamma < 1$ means $U_{stable} < U_{unstable}$, thus the system remains stable. This result implies that stability is determined by the square of the aspect ratio of tube wall *t/h* and swelling ratio $\lambda$. This also matches well with the result found in the literature ($\lambda_{cr} = 0.867 \cdot t^2/h^2$) [11]. $\lambda_{cr}$ required to trigger buckling instability is plotted as a function of the wall aspect ratio in Fig. 3(b), suggesting that slender walled tube becomes mechanically unstable at smaller swelling ratio.

Figure 4(a) is a stability map that can predict stability and buckling pattern together. With the horizontal and vertical axes representing **h/D** and **t/D**, respectively, any tube geometries can be mapped onto this plot. For given equilibrium swelling ratio $\lambda$, corresponding critical wall aspect ratio (**t/h**)$_{cr}$ for instability from (4) can be represented by a straight line drawn from the origin. The shaded area under this line is unstable region where $\gamma > 1$, hence samples fall into this region are expected to buckle. The slope of this boarder line increases with $\lambda$, making the unstable region larger. Furthermore, since buckling mode depends only on **h/D** as shown in (7), the buckling mode number can be determined based on the horizontal position of the sample on this map.

Collectively, stability of the swelling gel tube as well as buckling pattern can be predicted together from this plot.

To validate the theory, samples tested in swelling experiment in Fig. 1(b) are mapped onto the stability map in Fig 4(a). The critical stability line is drawn for $\lambda=1.5$. Samples on the same sloped line (**I-IV**) have the same instability index. Instability indices defined by (8) for each line are 0.25, 0.55, 2.18, and 4.98, respectively, which means that group **I** and **II** above the stability line should remain circular while group **III** and **IV** below the stability line are expected to create wrinkles. This prediction agrees with experimental result shown in Fig. 1(b). Discrepancy is found in only a few cases of **II-(i-iii)** (in the dotted circle) where nonlinear material behavior of gels at high stress is no longer negligible.

From (7), we know that samples aligned on the same vertical line (**i-vi**) should transform into patterns in the same buckling mode regardless of *t*. This was also experimentally observed in Fig. 1(b). Samples on the same column in Fig. 1(b) have the same *h*/*D* and their buckling modes are close to each other. The small difference across different groups should come from the thickness effect. For samples with thick wall, in-plane strain energy along the circumferential direction should also be considered, whereas this term is negligible for thin wall tube buckling where only out-of-plane strain energy along the circumferential direction is dominant. The experimental results for buckling mode numbers are plotted in Fig. 4(b). Instability patterns from samples spanning a wide range of physical dimension collapse well around theoretical prediction obtained from Fig. 3(a). This shows that we demonstrated full control over the pattern of gel tubes formed by mechanical instability. The pattern formation has also been simulated by finite element method (FEM), which adopts a coupled theory that considers the total free energy of the gel due both to the polymer network deformation and polymer-solvent mixing [24, 25]. The result also shows good agreement with FEM simulation. A set of representative results is shown in Fig. 1(b). See Supplemental Material [21] for detailed finite element analysis and full comparison between experimental and simulation result.

In summary, we have demonstrated well-controlled wrinkle formation of confined hydrogel tube using swelling-induced circumferential buckling. We have also developed a simple theory based on elastic energy and found that key dimensional parameters sensitive to stability and buckling pattern formation are thickness to height ratio and height to diameter ratio, respectively. Our experimental results showed good quantitative agreement with theoretical prediction as well as FEM simulation. In this study, it has been demonstrated that spontaneous formation of complex patterns can be achieved in a controlled manner by making use of mechanical instability of hydrogel. Furthermore, reversible nature of swelling and shrinking of hydrogel offers unique opportunities to develop versatile devices with tunable properties. We believe our study on buckling of swelling gels will contribute to increasing the breadth of possible application of soft materials in many emerging fields where complex morphologies and dramatic pattern shift are of critical importance, such as tissue engineering and tunable photonic/phononic band gap materials.

## ACKNOWLEDGMENTS

This work was supported by NSF and LLNL-LDRD grant. Authors would like to thank Dr. Kin Hung Fung for useful discussion. We also appreciate the Fulton High Performance Computing at Arizona State University to support our simulations.

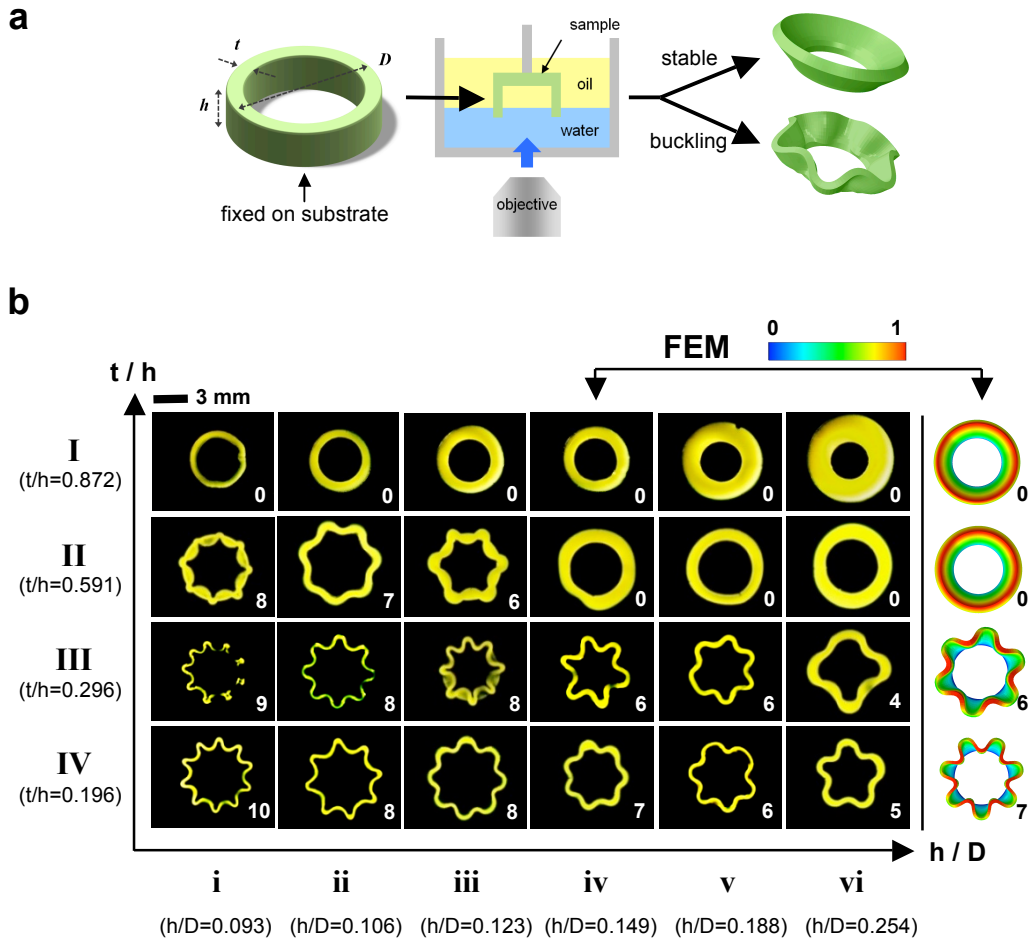

FIG. 1: (Color online) Pattern formation by circumferential buckling of gel tube (a) Characteristic dimensions of tubular gel and experimental setup. (b) Patterns formed in swelling experiment. Samples in the same row have the same *t/h*, and samples in the same column have the same *h/D* (Scale bar indicates 3 mm). Once buckles, the samples in the same row (i.e, with the same *h/D*) show the similar buckling patterns and the samples in the same column (i.e., with the same *t/h*) show the similar stability behavior. FEM simulation result for the group **iv** is also presented showing good agreement with experiment. (Color bar indicates normalized height.)

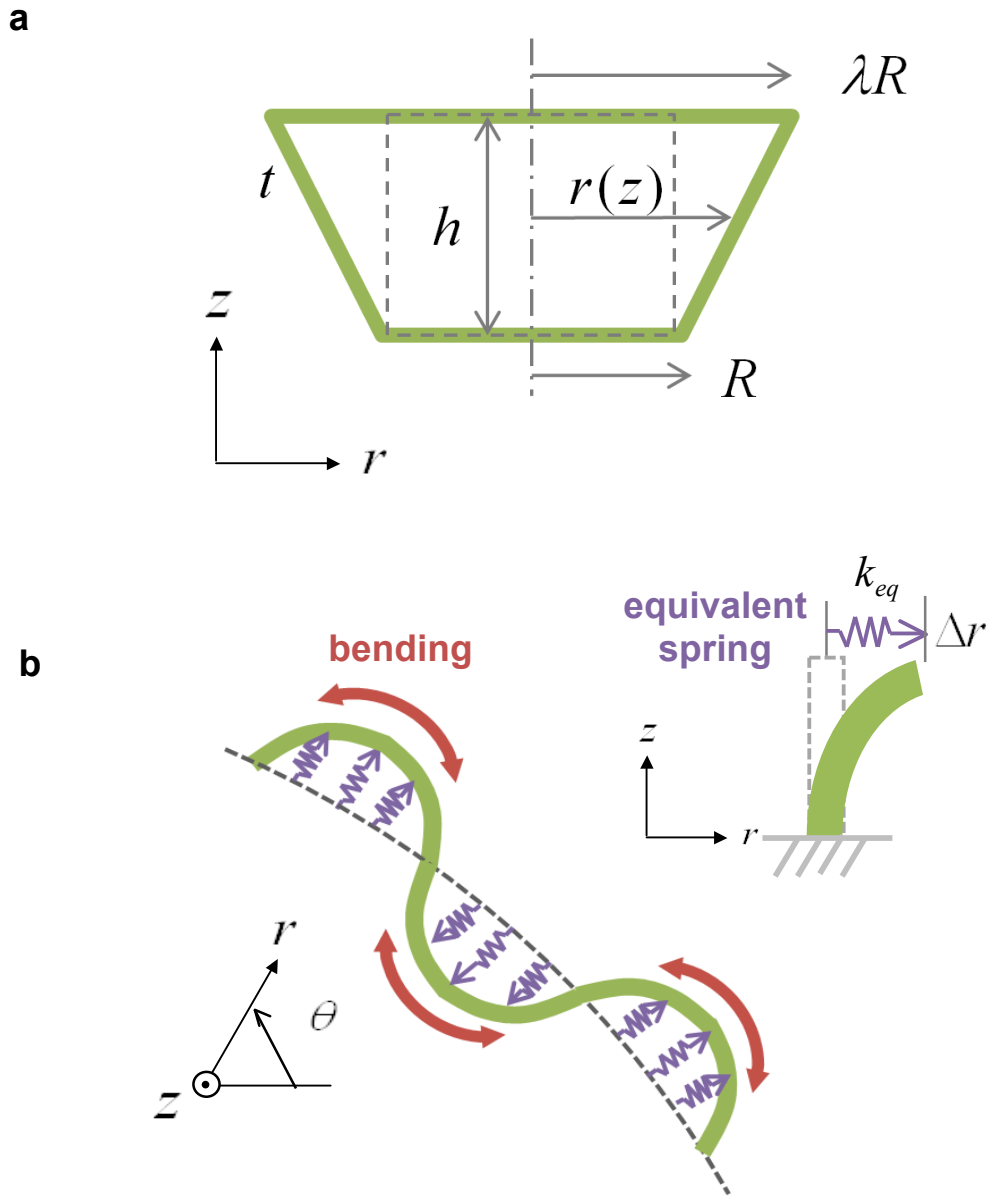

FIG. 2: (Color online) Model geometry (a) compressed configuration (side view) (b) buckled configuration (top view), deflection away from central axis modeled as equivalent springs

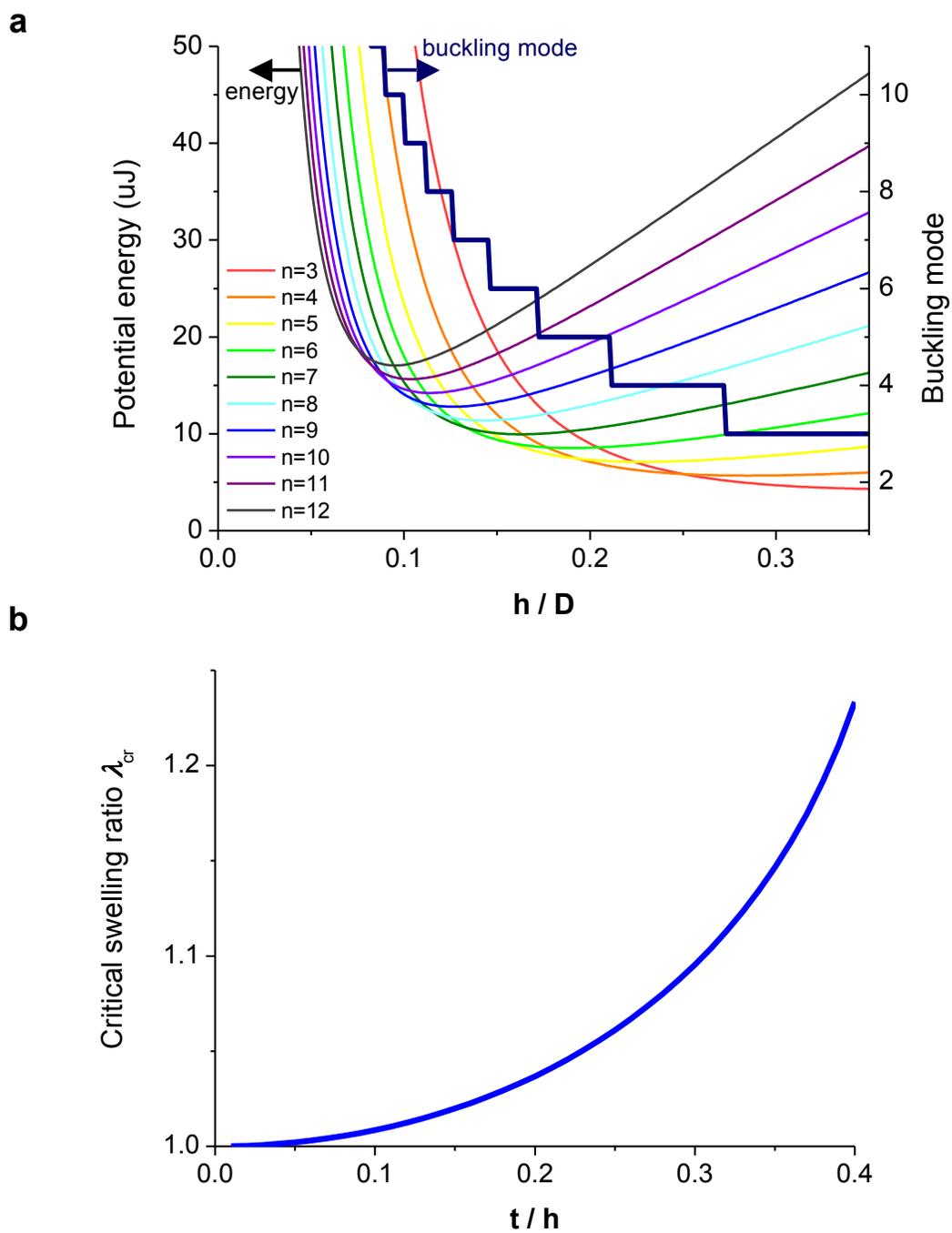

FIG. 3: (Color online) Theoretical prediction (a) Potential energy for different buckling mode. Minimum energy mode varies with *h/D*, resulting in different buckling patterns. (b) Critical swelling ratio for mechanical instability

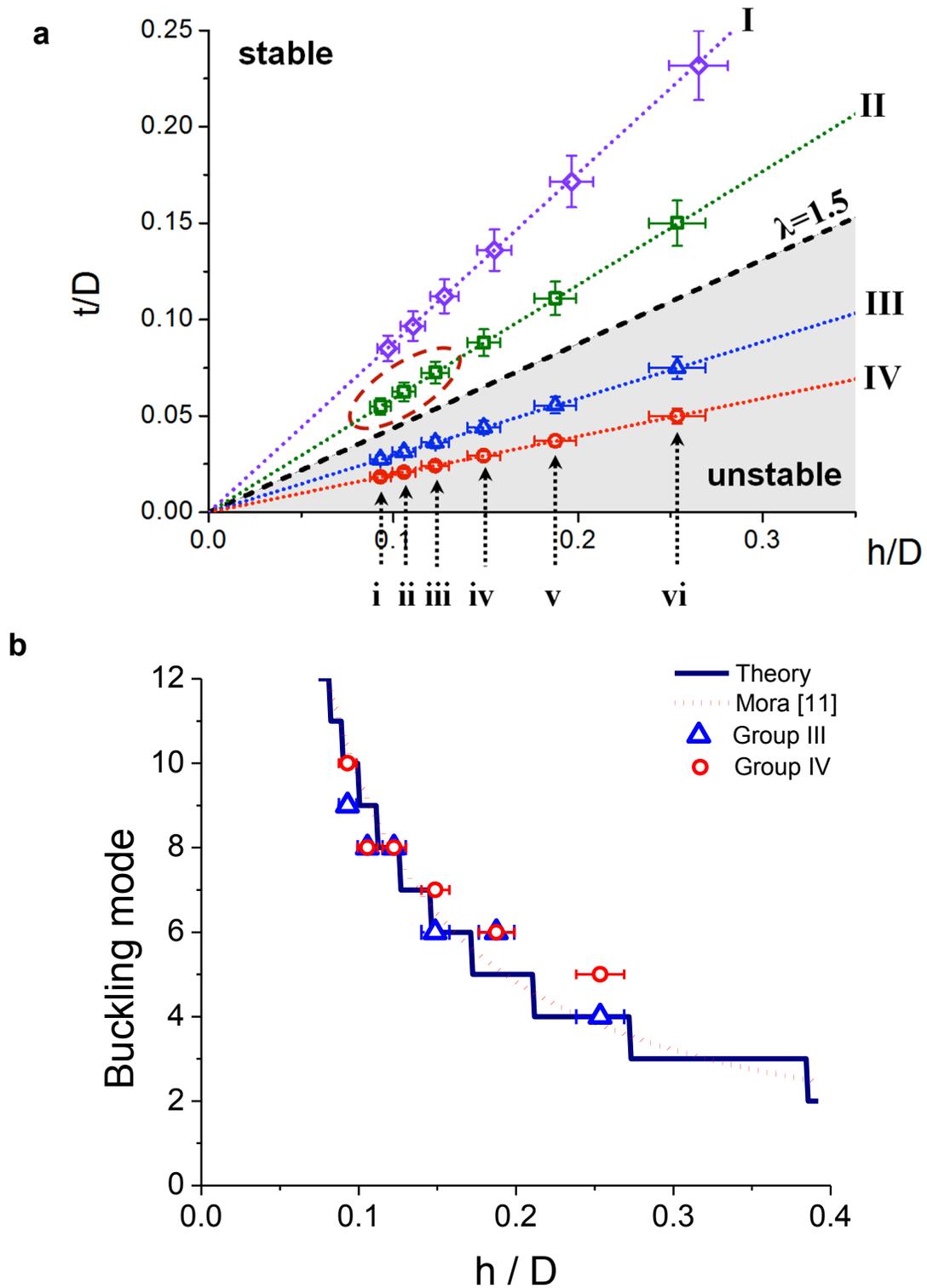

FIG. 4: (Color online) Stability and buckling pattern (a) Critical $t/h$ line (dotted) for $\lambda=1.5$ divides upper stable region and shaded unstable region below. Samples shown in Fig. 1b

are grouped into four groups (I-IV) by *t/h* for stability and into six groups (i-vi) by *h/D* for buckling pattern and are positioned on the stability map. (b) Buckling mode number from unstable samples from experiment. They match well with the prediction by present theory (solid line) as well as the analog of linear elasticity model for strip buckling from the literature (dotted line) [11].

# Supplemental Material for
# Prescribed Pattern Transformation in Swelling Gel Tubes by Elastic Instability

Howon Lee,[1] Jiaping Zhang,[2] Hanqing Jiang,[2] and Nicholas X. Fang[1,]*

[1]Department of Mechanical Engineering, Massachusetts Institute of Technology,

Cambridge, MA 02139, USA

[2]School for Engineering of Matter, Transport, and Energy, Arizona State University,

Tempe, AZ 85287, USA

## 1. Experimental Methods

### 1.1. Materials

Poly(ethylene glycol) diacrylate (PEGDA) hydrogel was used. Soft PEGDA was synthesized by mixing PEGDA (MW575, Sigma Aldrich) with PEG (MW200, Sigma Aldrich) in a weight ratio of 1:3, followed by addition of 0.5% (w/w) of photo-initiator (phenylbis(2,4,6-trimethylbenzoyl) phosphine oxide, Sigma Aldrich) for photopolymerization under ultraviolet (UV) illumination (wavelength: 405nm). Not being polymerized, PEG contributes to reducing crosslinking density by occupying intermolecular space between PEGDA during photo-polymerization, resulting in low modulus and large swelling ratio. Diameter of hydrogel disks in dry state and fully swollen state in de-ionized (DI) water was measured to obtain equilibrium swelling ratio $\lambda_{eq}$. Length-wise equilibrium swelling ratio of the hydrogel was 1.75. Elastic modulus of the hydrogel in swollen state measured by compression test was 0.11 MPa. This hydrogel dramatically changes optical property from transparent to opaque as it swells, which facilitates visualization of the shape of the swollen sample in experiment.

## 1.2. Fabrication

We used projection micro-stereolithography (PµSL) to fabricate 3D hydrogel tube samples. PµSL is a digital freeform microfabrication technology capable of fabricating complex shaped 3D micro architectures by using a dynamic mask generator and an UV light source coupled to a projection lens system to convert liquid monomer to solid polymer in an additive, layer-by-layer fashion [1]. Therefore, tube designs with different geometries and dimensions created by CAD software can be rapidly fabricated into actual samples. Physical dimensions of fabricated samples are listed in Table. S1. After fabrication, samples were rinsed in acetone for 2 hours to rinse uncured PEG out of the structure. Then samples were allowed to dry in a vacuumed desiccator for 30 minutes.

TABLE S1. Sample dimension

| Sample | | $D$ (µm) | $t$ (µm) | $h$ (µm) |
|---|---|---|---|---|
| I | i | 4188 ± 231 | 356 ± 20 | 406 ± 10 |
| | ii | 4188 ± 231 | 404 ± 22 | 463 ± 11 |
| | iii | 4188 ± 231 | 469 ± 26 | 535 ± 13 |
| | iv | 4188 ± 231 | 570 ± 31 | 648 ± 15 |
| | v | 4188 ± 231 | 718 ± 40 | 822 ± 19 |
| | vi | 4188 ± 231 | 971 ± 54 | 1110 ± 26 |
| II | i | 4653 ± 257 | 256 ± 14 | 433 ± 10 |
| | ii | 4653 ± 257 | 291 ± 16 | 492 ± 12 |
| | iii | 4653 ± 257 | 337 ± 19 | 570 ± 13 |
| | iv | 4653 ± 257 | 409 ± 23 | 692 ± 16 |
| | v | 4653 ± 257 | 516 ± 29 | 873 ± 21 |
| | vi | 4653 ± 257 | 698 ± 39 | 1180 ± 28 |
| III | i | 4653 ± 257 | 128 ± 7 | 433 ± 10 |
| | ii | 4653 ± 257 | 145 ± 8 | 492 ± 12 |
| | iii | 4653 ± 257 | 169 ± 9 | 570 ± 13 |
| | iv | 4653 ± 257 | 205 ± 11 | 692 ± 16 |
| | v | 4653 ± 257 | 258 ± 14 | 873 ± 21 |
| | vi | 4653 ± 257 | 349 ± 19 | 433 ± 28 |
| IV | i | 4653 ± 257 | 85 ± 5 | 492 ± 10 |
| | ii | 4653 ± 257 | 97 ± 5 | 570 ± 12 |
| | iii | 4653 ± 257 | 112 ± 6 | 692 ± 13 |
| | iv | 4653 ± 257 | 136 ± 8 | 873 ± 16 |
| | v | 4653 ± 257 | 172 ± 10 | 1180 ± 21 |
| | vi | 4653 ± 257 | 233 ± 13 | 433 ± 28 |

## 2. Modeling

The model system in our study is a cylindrical-walled hydrogel tube as shown in Fig. 1a in the main text. Diameter, height, and wall thickness of the tube are denoted as **D**, **h**, and **t**, respectively. Cylindrical coordinate system $(r,\theta,z)$ is used to describe deformation. Although there has been a report that bonding of gels to substrate plays an important role in buckling formation and mode [2], the interface between tubes and substrates have been carefully treated to achieve very strong bonding in our experiment. Therefore, base of the tube is constrained as a fixed boundary condition, whereas the rest of the structure is free to swell. When allowed to swell, there are two possible configurations for the swollen tube to adopt in order to accommodate expanded geometry in the original dimension; compression and buckling. The system chooses a shape that minimizes total potential energy. We assume that the entire structure first swells to the fully swollen state and then the bottom end is forced to fit into its original dry dimension. Therefore, fully swollen state is considered as stress-free reference state and any deformation from it has an elevated potential energy of the system. Although swelling of gel under constraints is not a purely elastic process involving a coupling of elastic and chemical equilibrium, this coupling effect does not play a significant role when the free energy of the compression state and the buckling state are compared because the free energy due to the coupling is assumed to be a constant for both configurations. This permits pure elastic consideration. The elastic energy for each configuration is analyzed to predict stability as well as post-buckling pattern.

## 2.1. Stable configuration

For stable configuration, cross-section of the tube remains circular with the radius being a function of height only. It is assumed that radius is linearly varying from the dry radius $R(=D/2)$ at the fixed bottom to the fully swollen radius $\lambda R$ at the top as shown in Fig. S1a, where $\lambda$ denotes stretch ratio. Then, radius can be written as

$$r = r(z) = R\left[1 + (\lambda - 1)\cdot \frac{z}{h}\right] \tag{1}$$

Since only in-plane compression is involved, total elastic energy in the stable configuration is obtained as

$$U_{stable} = \int \frac{1}{2} E \varepsilon_{\theta\theta}^2 dV = \frac{1}{24}\pi EDth \cdot b(\lambda) \tag{2}$$

where $\varepsilon_{\theta\theta}(z) = \dfrac{r - \lambda R}{\lambda R} = -(1-\dfrac{1}{\lambda})(1-\dfrac{z}{h})$ is strain in circumferential direction and $b(\lambda) = (1 - 1/\lambda)^2 (3 + \lambda)$.

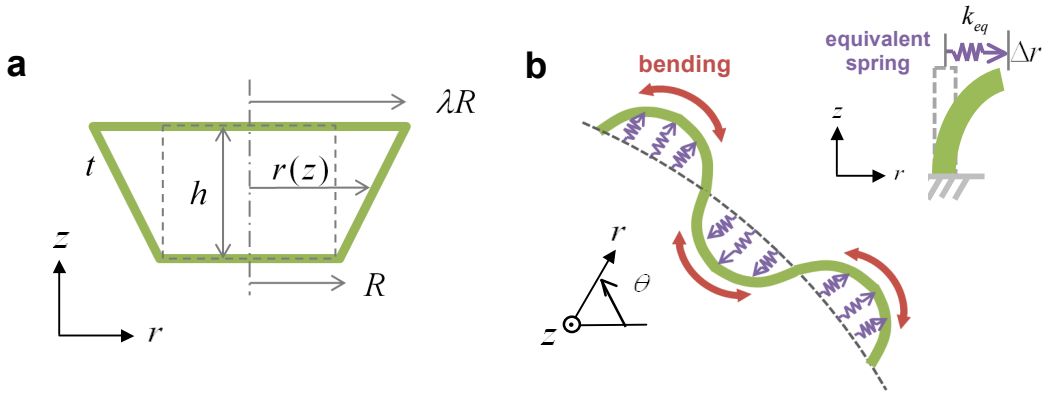

FIG. S1. Model geometry (a) compressed configuration (side view) (b) buckled configuration (top view), deflection away from central axis modeled as equivalent springs

## 2.2. Buckled configuration

Once the structure becomes mechanically unstable, it buckles and creates wrinkles along its circumference. Due to the confinement of the tube at its bottom surface, the gel tube swells more near its upper end than near its bottom end. Two parts of energy should be considered in this case. The first part is the elastic energy due to the wavy bending along the circumferential direction. This energy contribution increases with buckling mode because the wall undergoes more bending with large curvature in higher buckling mode. This part of energy is referred to circumferential energy. The second part is the elastic energy due to the deflection of the wall in axial direction. As the gel swells more near its free upper end than near its confined bottom, the gel wall has to deflect outwards or inwards in axial direction depending on the position on the wave. This energy contribution decreases with buckling mode because higher buckling mode results in smaller wave amplitude in given length, thus less deflection in axial direction is necessary. This part of energy is referred to axial energy. It should be noted that the name of energy is purely for purpose of analysis and simplification. With the two energy contributions working together, there exists an optimum buckling mode that yields minimum total potential energy.

### 2.2.1. Derivation of the closed form wave amplitude

In order to describe the wavy pattern of buckled configuration in an analytical form, one should be able to express the wave amplitude with known variables. Assuming that wavy pattern follows sinusoidal function along the circumference, amplitude of sinusoidal wave for given contour length can be calculated and obtained in a closed form using approximation for the elliptic integral [3]. Given overall wave contour length, wave amplitude is inversely proportional to the number of waves as shown is Fig. S2. First, we write radius on the top end of the wrinkled tube as

$$r(\theta, z = h) = R + A\cos n\theta \tag{3}$$

where $A$ is amplitude of the wave to be determined and $n$ is mode number, i.e. the number of waves along the circumference. The contour length of the waves could be integrated as

$$L = 4n \int_0^{\pi/2n} \sqrt{dr^2 + (rd\theta)^2} \tag{4}$$

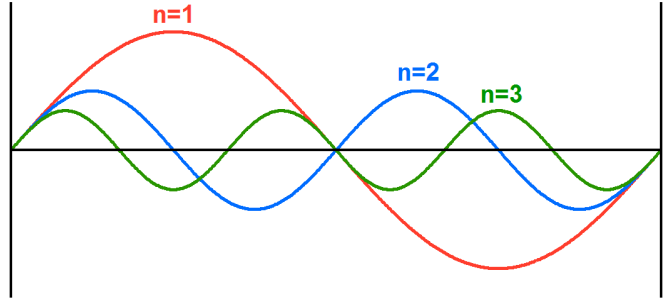

FIG. S2. Given arc-length of sinusoidal wave function, wave amplitude becomes smaller as wave number increases.

Substituting (3) into (4) under small strain assumption, and using the approximation for elliptic integral given by [3]

$$\int_0^{\pi/2} \sqrt{1 + k^2 \cos^2 x}\, dx \approx \frac{\pi}{2}\left(3 - \frac{2}{\sqrt{1 + k^2/4}}\right) = L \tag{4}$$

The arc-length of the wave can be obtained by

$$\begin{aligned} L &= 4n \int_0^{\pi/2n} \sqrt{1 + \left(\frac{nA}{R}\right)^2 \sin^2 n\theta}\, R\, d\theta \\ &= 4R \cdot \left[\frac{\pi}{2}\left(3 - \frac{2}{\sqrt{1 + (nA/R)^2/4}}\right)\right] \end{aligned} \tag{5}$$

From stable buckling, the contour length is

$$L = 2\pi\lambda R \tag{6}$$

Combine eqns. (5) and (6), solve for $A$ yields

$$A = \frac{D}{n}a(\lambda) \tag{7}$$

where $a(\lambda) = \sqrt{\left(\frac{2}{3-\lambda}\right)^2 - 1}$.

Then wrinkled radius at the free top end (i.e. $z = h$) can be now expressed with given parameters as follows.

$$r(\theta, z = h) = R\left\{1 + \frac{2}{n}a(\lambda)\cos n\theta\right\} \tag{8}$$

### 2.2.2. Elastic energy for buckled configuration

Radial wave amplitude varies with height and here we assume that it follows a bending profile of cantilever beam subjected to point load at the free end, $z^2(3h-z)/(2h^3)$, where $h$ is the height of the cantilever beam and $z$ is the specific location along the height direction. Then, radius of wrinkled cross-section can be written as

$$r(\theta, z) = R\left\{1 + \frac{2}{n}a(\lambda)\cos n\theta \cdot \frac{z^2(3h-z)}{2h^3}\right\} \tag{9}$$

Elastic energy from bending along the circumferential direction is obtained by [4]

$$U_c = \frac{1}{2}\int EI\kappa^2 dV = \frac{11}{140}\frac{\pi E t^3 a^2(\lambda)h}{D}n^2 \tag{10}$$

where $I = \dfrac{t^3 R d\theta}{12}$ and $\kappa \approx \dfrac{1}{R^2}\dfrac{\partial^2 r(\theta,z)}{\partial \theta^2}$ are bending moment of inertia and curvature of the wave. Note that energy is proportional to $n^2$, which means that lower mode is energetically favorable.

Next, elastic energy along the axial direction is considered. This part of energy is modeled as bending of a set of cantilever beams surrounding the central axis. For simplicity, equivalent springs for cantilever beams are introduced as shown in Fig. S1b. Then each spring undergoes stretching by the distance to the neutral circumferential line, which is the wave amplitude at each point, $\Delta r = r(\theta; z = h) - R$. From beam theory, equivalent spring constant of each cantilever beam is given by [5]

$$k_{eq} = \dfrac{Et^3 R d\theta}{4h^3} \qquad (11)$$

Energy for the wall is obtained by integration of energy of each individual cantilever beam:

$$U_a = \int \dfrac{1}{2} k_{eq} \left(\Delta r\right)^2 = \dfrac{\pi E t^3 D^3 a^2(\lambda)}{16 h^3} \cdot \dfrac{1}{n^2} \qquad (12)$$

Note that the energy in this case is inversely proportional to $n^2$, which means that higher mode is energetically favorable. Combining Eq. (10) and Eq. (12), total elastic energy for buckled configuration, therefore, is given by

$$U_{unstable} = U_c + U_a = \dfrac{\pi E t^3 h a^2(\lambda)}{D}\left[\dfrac{11}{140} n^2 + \dfrac{1}{16(h/D)^4}\cdot\dfrac{1}{n^2}\right] \qquad (13)$$

Minimization of $U_{unstable}$ will give an optimum mode number $n$ for buckling.

## 3. Coupled field theory and finite element implementation

Consider a hydrogel occupy volume $V$ and enclosed by boundary $S$. The gel is subjected to body force $B_i$ and surface traction $T_i$. When contact with solvent, there are $i$ amount of solvent passing through the interface per unit time. The free energy density change of the whole thermodynamic system $\delta G$ includes the free energy density change of the gel $\delta W$ and work done by the external mechanical load and chemical load.

$$\frac{\delta G}{\delta t} = \int \frac{\delta W}{\delta t} dV - \int B_i \frac{\delta x_i}{\delta t} dV - \int T_i \frac{\delta x_i}{\delta t} dA - \int \mu i dA \tag{14}$$

The free energy of the gel $W$ depends both on deformation characterized by deformation gradient $\mathbf{F} = \frac{\partial \mathbf{x}}{\partial \mathbf{X}}$ and solvent flux characterized by solvent concentration $C$, i.e., $W = W(\mathbf{F}, C)$. Thermodynamics indicates that the free energy of the whole system should not increase, i.e., $\frac{\partial G}{\partial t} \leq 0$. The equality holds if there is no energy dissipation. Due to fast local rearrangement of molecules, it is reasonable to assume that the energy dissipated due to viscosity is very small and negligible. Consequently, it leads to two constitutive relations expressed by nominal stress $\mathbf{s}$ and chemical potential $\mu$,

$$s_{iK} = \frac{\partial W(\mathbf{F}, C)}{\partial F_{iK}}, \mu = \frac{\partial W(\mathbf{F}, C)}{\partial C} \tag{15}$$

Another simplification is the polymer network of the gel be incompressible [6]. This statement means that the volumetric change of the gel is purely caused by the amount of solvent molecule diffuse in or out. The solvent concentration is thus related to deformation by

$$1 + vC = \det(\mathbf{F}) \tag{16}$$

This condition could be tacitly enforced by introducing another free energy $U$ using Legendre transformation

$$U = W - \mu C \tag{17}$$

Consequently, the nominal stress $s_{iK}$ now depends on $U = U(\mathbf{F}, \mu)$ via $s_{iK} = \partial U(\mathbf{F}, \mu) / \partial F_{iK}$. Depending on the materials, $W$ could have many different forms. A specific form of $W$ would be

$$\begin{aligned} W &= \frac{1}{2} N k_B T \left[ \lambda_1^2 + \lambda_2^2 + \lambda_3^2 - 3 - 2\ln(\lambda_1 \lambda_2 \lambda_3) \right] \\ &- \frac{k_B T}{v} \left[ vC \log\left(1 + \frac{1}{vC}\right) + \frac{\chi}{1+vC} \right] \end{aligned} \tag{18}$$

where the first term represents deformation energy of the polymer network [7] and the second term represents mixing energy [8] between the polymer and solvent. $N$ is number of polymer chains per volume hydrogel. $k_B$ is Boltzmann constant and $T$ is temperature. $\lambda_1, \lambda_2$ and $\lambda_3$ are principal stretch values. $v$ is volume of solvent molecules. $\chi$ is the polymer solvent interactive parameter.

Typically, the gel has some initial swelling because it is exposed to environment. In this case, assume an initial homogeneous state where $\lambda_1 = \lambda_2 = \lambda_3 = \lambda_0$. The previous stretch $\lambda$ changed to $\lambda \lambda_0$ and determinant of deformation gradient $\det(\mathbf{F})$ changed to $\lambda_0^3 \det(\mathbf{F})$. In addition, combining Eq. (16) and Eq. (18), Eq. (17) is simplified to

$$\begin{aligned} U &= \frac{1}{2} N k_B T \left[ \left( \lambda_1^2 + \lambda_2^2 + \lambda_3^2 \right) \lambda_0^2 - 3 - 2\ln\left( \lambda_0^3 \det(\mathbf{F}) \right) \right] \\ &- \frac{k_B T}{v} \left[ \left( \lambda_0^3 \det(\mathbf{F}) - 1 \right) \ln \frac{\lambda_0^3 \det(\mathbf{F})}{\lambda_0^3 \det(\mathbf{F}) - 1} + \frac{\chi}{\lambda_0^3 \det(\mathbf{F})} \right] - \frac{\mu}{v} \left[ \lambda_0^3 \det(\mathbf{F}) - 1 \right] \end{aligned} \tag{19}$$

Cauchy stress $\boldsymbol{\sigma}$ is a direct stress measure because it is defined by the deformed area and volume. It relates to nominal stress $\mathbf{s}$ by

$$\sigma_{ij} = \frac{1}{\det(\mathbf{F})} s_{iK} F_{jK} \tag{20}$$

Combining Eq. (15), Eq. (17) and Eq. (19), we have

$$\sigma_{ij} = Nk_B T \lambda_0^2 \left[\det(\mathbf{F})\right]^{-\frac{1}{3}} \bar{B}_{ij} + \frac{k_B T}{v} \delta_{ij} \left[ -\frac{Nv}{\det(\mathbf{F})} - \lambda_0^3 \ln\left(\frac{\det(\mathbf{F})\lambda_0^3}{\det(\mathbf{F})\lambda_0^3 - 1}\right) \right. \\ \left. + \frac{1}{\det(\mathbf{F})} + \frac{\chi}{\det(\mathbf{F})^2 \lambda_0^3} - \frac{\mu}{k_B T} \lambda_0^3 \right] \tag{21}$$

In the above equation, $\bar{B}_{ij} = \det(\mathbf{F})^{-\frac{2}{3}} F_{iK} F_{jK}$ is the normalized left Cauchy-Green tensor. Taking variation of the stress and adopting a similar procedure as in Kang [9], the tangential stiffness could be obtained

$$C_{ijkl} = Nk_B T \lambda_0^2 \det(\mathbf{F})^{-\frac{1}{3}} H_{ijkl} + \frac{k_B T}{v} \left( -\frac{\chi}{\det(\mathbf{F})^2 \lambda_0^3} + \frac{\lambda_0^3}{\det(\mathbf{F})\lambda_0^3 - 1} \right. \\ \left. - \lambda_0^3 \ln\frac{\det(\mathbf{F})\lambda_0^3}{\det(\mathbf{F})\lambda_0^3 - 1} - \frac{\mu}{k_B T} \lambda_0^3 \right) \delta_{ij}\delta_{kl} \tag{22}$$

In the above equation,

$$H_{ijkl} = \frac{1}{2}\left(\bar{B}_{jl}\delta_{ik} + \bar{B}_{ik}\delta_{jl} + \bar{B}_{jk}\delta_{il} + \bar{B}_{il}\delta_{jk}\right) \tag{23}$$

We implement above formulation in a commercial finite element package ABACUS (Abacus Solutions, LLC.). Equations (21) and (22) share similar form as equations (44) and (53) in Kang [9], they are indispensable components to write a user material subroutine (UMAT) in a ABAQUS. Both UMAT and a in-house program updated from an earlier work [10] are capable to solve this equilibrium swelling problem. For simplicity, UMAT is used here. In the UMAT, the gel is programmed as a user material and chemical potential is an adjustable parameter. Swelling is realized via coupled-temperature displacement analysis by changing chemical potential. To simulate the buckling process, an initial perturbation analysis is performed to generate the buckling wave followed by a subsequent post-buckling analysis to produce the final wave shape. 0.1% of the first three buckling modes magnitude from a thermal expansion perturbation analysis (expansion ratio=1.75) is added as an initial geometric imperfection for the post-buckling analysis. Different small initial imperfection is tested and exhibits no significant influence over the post-buckling result. This agrees well with the general understanding that initial geometric imperfection is merely a "trigger" for post-buckling study. Parameters used in the simulation are: $k_B T/v = 138$ MPa for water; $Nv = 3.06 \times 10^{-4}$ corresponds to gel elastic modulus of 0.11 MPa, $\chi = 0.57$; degree of swelling is adjusted (based on deformation estimation) by increasing chemical potential gradually, from $\mu/kT = -2.02$ ($\lambda_0 = 1.01$) to $\mu/kT = 0$ (corresponding to $\lambda = 1.75\lambda_0$) while maintaining $Nv$ and $\chi$ the same. Typically 1000~3000 incremental steps in ABAQUS are required to complete a single simulation due to the strong divergence. Several thousands of 3-D brick elements with temperature degree of freedom (C3D8T) are used for most of the calculations and result reaches convergence by mesh refinement. FEM simulation results are presented in Fig. S3 and they are in very good agreement with experimental results.

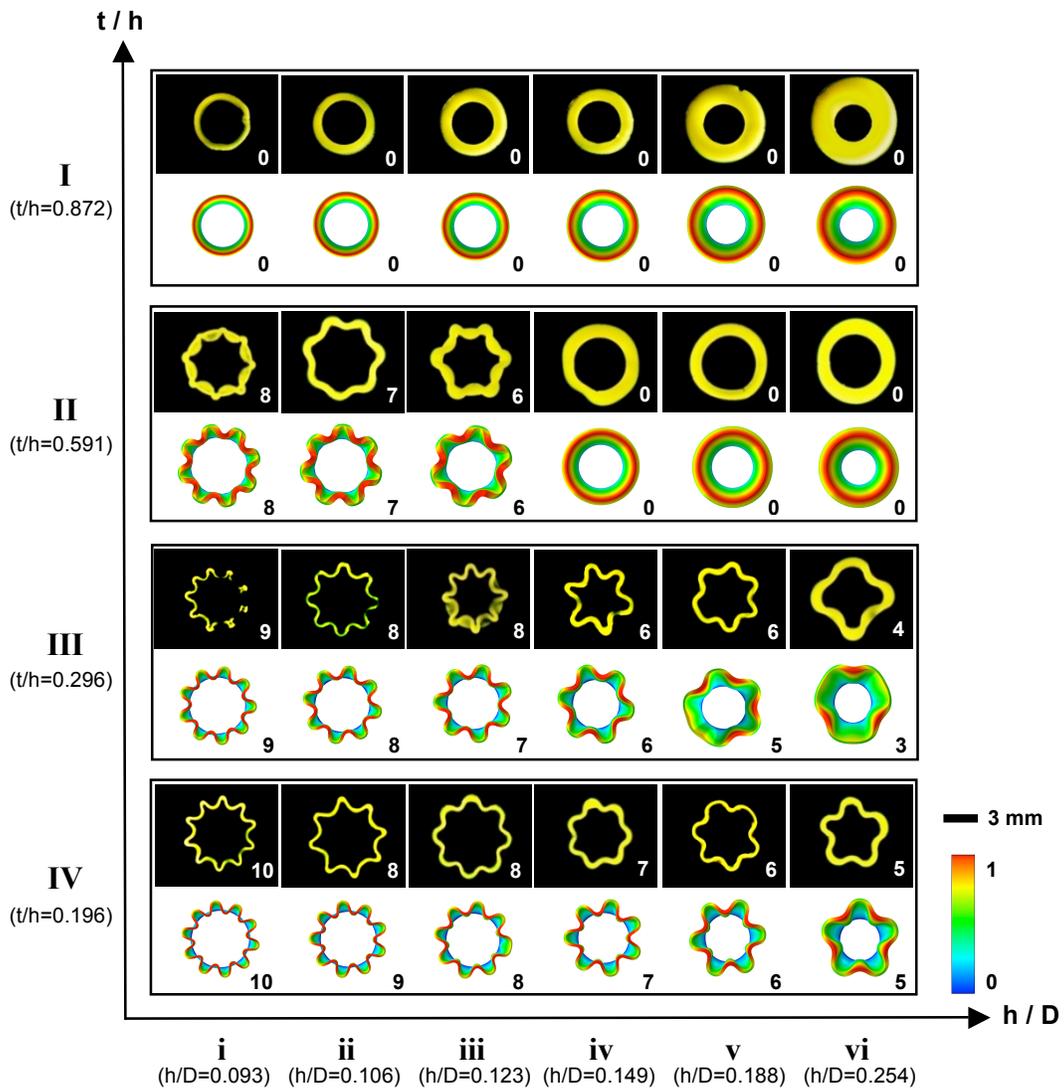

FIG. S3. Comparison between FE result and experimental result. Color bar indicates normalized height of samples. Scale bar indicates 3mm.

**Supplemental Movie S1.** A gel tube which remains in stable configuration during swelling experiment. The movie plays 50 times faster than real time. (MovieS1.avi; 5.5 MB).

**Supplemental Movie S2.** A gel tube which transforms into buckled configuration with 5 lobes along the circumference during swelling experiment. The movie plays 50 times faster than real time. (MovieS2.avi; 4.5 MB).